# Effect of mixed pinning landscapes produced by 6 MeV Oxygen irradiation on the resulting critical current densities $J_c$ in 1.3 μm thick GdBa$_2$Cu$_3$O$_{7-d}$ coated conductors grown by co-evaporation

N. Haberkorn,[1,2] S. Suárez,[1,2] P. D. Pérez,[1] H. Troiani,[1,2] P. Granell,[3] F. Golmar,[3,4] Jae-Hun Lee,[5] S. H. Moon.[5]

[1] *Consejo Nacional de Investigaciones Científicas y Técnicas, Centro Atómico Bariloche, Av. Bustillo 9500, 8400 San Carlos de Bariloche, Argentina.*

[2] *Instituto Balseiro, Universidad Nacional de Cuyo, Av. Bustillo 9500, 8400 San Carlos de Bariloche, Argentina*

[3] *INTI, CMNB, Av. Gral Paz 5445 (B1650KNA), San Martín, Buenos Aires, Argentina.*

[4] *Consejo Nacional de Investigaciones Científicas y Técnicas, Escuela de Ciencia y Tecnología, UNSAM, Campus Miguelete, (1650), San Martín, Buenos Aires, Argentina.*

[5] SuNAM Co. Ltd, Ansung, Gyeonggi-Do 430-817, South Korea.

*e-mail* corresponding author: *nhaberk@cab.cnea.gov.ar* Tel : +54 0294 444 5147- FAX: +54 0294 444 5299

## Abstract

We report the influence of crystalline defects introduced by 6 MeV $^{16}$O$^{3+}$ irradiation on the critical current densities $J_c$ and flux creep rates in 1.3 μm thick GdBa$_2$Cu$_3$O$_{7-\delta}$ coated conductor produced by co-evaporation. Pristine films with pinning produced mainly by random nanoparticles with diameter close to 50 nm were irradiated with doses between $2\times10^{13}$ cm$^{-2}$ and $4\times10^{14}$ cm$^{-2}$. The irradiations were performed with the ion beam perpendicular to the surface of the samples. The $J_c$ and the flux creep rates were analyzed for two magnetic field configurations: magnetic field applied parallel (**H**∥*c*) and at 45° (**H**∥45°) to the *c*-axis. The results show that at temperatures below 40 K the in-field $J_c$ dependences can be significantly improved by irradiation. For doses of $1\times10^{14}$ cm$^{-2}$ the $J_c$ values at μ$_0$H = 5 T are doubled without affecting significantly the $J_c$ at small fields. Analyzing the flux creep rates as function of the temperature in both magnetic field configurations, it can be observed that the irradiation suppresses the peak associated with double-kink relaxation and increases the flux creep rates at intermediate and high temperatures. Under 0.5 T, the flux relaxation for **H**∥*c* and **H**∥45° in pristine films presents characteristic glassy exponents $\mu$ = 1.63 and $\mu$ = 1.45, respectively. For samples irradiated with $1\times10^{14}\,cm^{-2}$, these values drop to $\mu$ = 1.45 and $\mu$ =1.24, respectively

**1. Introduction**

During the last years, a great amount of effort has been orientated to improve the superconducting properties of $R$Ba$_2$Cu$_3$O$_{7-\delta}$ ($R$BCO; $R$: Sm, Dy, Y, Gd) coated superconductors (CCs) [1,2,3,4]. One of the strongest aspects expected for applications such as motors and magnets is to be able to maintain high critical current density $J_c$ at high magnetic fields [5,6]. The current carrying capacity is determined by the interaction of vortex matter with crystalline defects [7]. The geometry and density of the pinning centers determine the in-field and angular dependence of $J_c$ [7,8,9]. In addition, the vortex-defect interactions are not accumulative and the vortex pinning produced by the different crystalline defects depends on the temperature and on the magnetic field. While small defects improve pinning mainly below 40 K, large defects (size of a few coherence length $\xi$) do so throughout the whole temperature range. [10]. For the latter, the pinning can be isotropic (nanoparticles) or correlated (columnar defects). Innovative approaches have been used to design optimal vortex pinning by defect structures [8,11]. Most of them are based on the inclusion of nanometric secondary phases [12,13].

The formation of the desired pinning centers depends sensitively on the film deposition technique and substrate architecture [1,12,13,14]. Despite the remarkable advance in the synthesis of desirable microstructures obtained at laboratory scale research, pinning landscapes including a high density of strong and weak pinning centers are usually difficult at industrial scale. This limitation in the synthesis of CCs has been recently overcome by including pinning centers by irradiation [15,16,17,18]. Depending on the mass and energy of the ions and the properties of the superconducting material, irradiation enables the creation of defects with well-controlled density and topology such as points, clusters, or tracks [19]. The in-field dependence of $J_c$ at temperatures below 40 K has been improved by irradiation with protons in CCs grown by metal organic deposition MOD and co-evaporation [15,16]. Most recently, the feasibility to similarly improve the properties of CCs obtained by MOD in much shorter irradiation times using 3.5 MeV $^{16}O^{3+}$ has been reported [17].

Here, we show the effects of random point defects and nanoclusters introduced by 6 MeV $^{16}O^{3+}$ irradiation with doses between $2 \times 10^{13}$ cm$^{-2}$ and $4 \times 10^{14}$ cm$^{-2}$ on the magnetic field dependence of $J_c$ 1.3 µm thick GBCO thin films grown by co-evaporation. In addition, the vortex dynamics displayed in pristine and irradiated samples is analyzed by performing magnetic relaxation measurements. The microstructure of the pristine tape displays random and irregular shape precipitates with typical size of 50 nm. Unlike ref. [17], the energy of the ions was selected to place the Bragg peak inside the substrate (instead of the superconducting layer). Several attempts of irradiation with 3.5 MeV $^{16}O^{3+}$ showed that, although the in-field of $J_c$ dependence is smoother than for pristine samples, the irradiation reduces considerably the low field $J_c$ values. In comparison to our previous work [16], in which tapes irradiated with protons reached the optimal doses in tens of minutes, the irradiation with 6 MeV $^{16}O^{3+}$ demands only a few seconds. While the irradiation significantly improves the in-field dependence of $J_c$ for the **H** ∥ *c* and **H**∥45° configurations, its absolute values are merely affected at low fields. From Maley analyses, we calculate the µ glassy exponent for the collective creep regime. For both configurations, the µ values decrease after irradiation, which increases the flux creep rates in comparison to pristine films.

**2. Material and methods**

The GBCO tape was grown by the co-evaporation technique previously described in refs. [4,20]. The magnetization (**M**) measurements were performed by using a superconducting quantum interference device (SQUID) magnetometer with the applied magnetic field (**H**) parallel to the *c*-axis (**H**∥c) and rotated 45° from the c axis (**H** ∥ 45°). The $T_c$ values used in this work (based on magnetization data) were determined from *M* (*T*) at $\mu_0 H$ = 0.5 mT applied after zero field cooling. The $J_c$ values were calculated from the magnetization data using the appropriate geometrical factor in the Bean Model. For **H**∥c, $J_c = \frac{20 \Delta M}{w(1 - w/3l)}$, where *ΔM* is the difference in magnetization between the top and bottom branches of the hysteresis loop, *t* is the thickness, and *l* and *w* are the length and width of the film ($l > w$), respectively. For **H** ∥ 45°, both the longitudinal ($M_l$, parallel to **H**) and transversal ($M_t$, perpendicular to **H**) components of **M** were measured. Then the irreversible magnetization was calculated as $\Delta M = (\Delta M_l^2 + \Delta M_t^2)^{1/2}$. Using the anisotropic version of the critical state Bean model was obtained $J_c = \dfrac{20\Delta M}{w\left(1 - \dfrac{w\cos(45°)}{3l}\right)}$ [21]. The creep rate measurements were recorded for more than 1 hour. The initial time was adjusted considering the best correlation factor in the log-log fitting of the $J_c$ ($t$) dependence. The initial critical state for each creep measurement was generated using $\Delta H \sim 4H^*$, where $H^*$ is the field for full-flux penetration [22]. A good correlation between $J_c$ values obtained by magnetization and by electrical transport in coated conductors at low temperatures has been reported [15].

Irradiation with 6 MeV Oxygen (O-irradiation) produces random point defects and nm-sized anisotropic defects [17,19]. The cumulative amount of displacement damage (displacements per atom, DPA) after each dose (as estimated using the SRIM code [23]), varies from 0.02 up to 0.8 DPA within the sample and for the doses used in the present work. The irradiation was performed at room temperature on pieces with typical area 1.5 x 1.5 mm using ion beam currents of ≈ 4 nA. In all the cases, the ion beam was applied perpendicular to the surface of the samples. To guarantee good thermal contact, the samples the pieces were fixed to the holder with silver paint. The irradiation spot was 2 mm in diameter, and the homogeneity of the beam was tested impacting in a fluorescent sample ($F_2Ca$, $AlF_3$ or LiF), which allowed to take a direct measure of the beam spot dimensions. The irradiations were performed with the ion beam positioned at the center of the sample. Considering the beam spot dimension and the sample geometry (square with size 1.5 x 1.5 mm$^2$), the corner of the samples have probably not been homogenously irradiated. All the samples display similar properties ($T_c$ and $J_c$ ($H$, $T$)) before irradiation. Wherever used, the notation IRRx indicates a GBCO film without irradiation (x=0); and x= 2, 4, 6, 10 and 40 correspond to films irradiated with oxygen dose 2x10$^{13}$ cm$^{-2}$, 4x10$^{13}$ cm$^{-2}$, 6x10$^{13}$ cm$^{-2}$, 1x10$^{14}$ cm$^{-2}$ and 4x10$^{14}$ cm$^{-2}$, respectively.

## 3. Results and discussion

The pinning landscape of pristine films was analyzed from cross-sectional TEM images (Fig. 1*a*). The microstructure displays sphere-like and irregular precipitates embedded in the GBCO matrix with typical diameter of approximately 50 nm (Fig. 1*b*). EDS analysis indicates that the precipitates can be associated with $Gd_2O_3$. Sphere-like nanoparticles produce isotropic pinning [24]. Irregular precipitates are extended mainly along *ab*- plane and the *c*-axis. Nanoparticles aligned at the *c*-axis are expected to assist correlated pinning usually present in thin films due to defects such as dislocations and twin boundaries [13]. The precipitates are typically spaced between ~100 nm and 200 nm, which corresponds to an equivalent matching field < 0.2 T. This is in agreement with high $J_c$ values at low fields and poor in-field dependences [16]. The inclusion of defects by oxygen irradiation in pristine CCS was analyzed in ref. [17]. Similar to proton irradiation, oxygen irradiation creates random point disorder and a large number of finely dispersed small defects (ø 5 nm approximately) [15,17].

The $T_c$ in the as-grown film is 92.7 K (see Fig. 2) and is gradually suppressed with the doses. Table 1 shows the $T_c$ values after irradiation. The irradiated samples display a small kink close to zero magnetization, which can be associated with inhomogeneous irradiation at the corners. In addition, irradiated samples display an extended magnetic flux penetration close to $T_c$. This fact can be associated mainly with weaker vortex pinning due to an increment in the vortex fluctuations close to $T_c$ [25]. The structural disorder produced by the irradiation should increases the penetration length ($\lambda$) [26, 27]. Also, a slight increment in $\lambda$ should affect the transition width $\Delta T_c$. Thermal fluctuations broaden the superconducting transition by $\Delta T_c \geq G_i * T_c$ [25], where *Gi* is the Ginzburg number $G_i = \frac{1}{2}\left[\frac{\gamma T_c}{H_c^2(0)\xi^3(0)}\right]^2$ (with $H_c$ the thermodynamic critical field, $\gamma$ the mass anisotropy and $\xi$ the coherence length). In optimal doped YBCO ($\lambda$ = 140 nm, $T_c$ = 92 K, $\gamma$ = 5 and $\xi$ = 1.2 nm), Gi ≈ 0.018.

Due to the fact that random disorder and nanoclusters originated by proton and oxygen irradiation produce the most significant changes in $J_c$ at low temperatures [15,16,17], our study focuses on $J_c$ (*H*) at temperatures below 40 K. The irradiation for doses larger than IRR2 affects both $J_c$ (H) and the vortex dynamics of the samples. Figures 3*a-c* shows a comparison between $J_c$ (H) for IRR0 and IRR10 at 5 K, 27 K and 40 K. The curves display two different regimes. At low fields (regime (I)), $J_c$ is almost constant up to a characteristic crossover field $H^*$ (usually called $B^*$ [28]). Then, it is followed by a power-law regime ($J_c \propto H^{-\alpha}$) at intermediate fields (regime II), observed as a linear dependence on the *log–log* plot. The main changes produced by the irradiation are manifested at high magnetic fields and can be related to a reduction in the $\alpha$ values. At $\mu_0 H = 5$ T, IRR10 displays $J_c$ values between 1.7 and 2 times larger than IRR0. Figures 4*a-c* show the doses dependence of $\alpha$, $J_c$ ($\mu H_0 = 0$) and $J_c$ ($\mu H_0 = 3$ T) at 5 K, 27 K and 40 K. The increment in the doses from $2 \times 10^{13}$ cm$^{-2}$ to $1 \times 10^{14}$ cm$^{-2}$ reduces systematically the $\alpha$ value for the three analyzed temperatures without appreciably affecting the $J_c$ values at low fields (Figs 4*ab*). For example, at 5 K (relevant for magnets [6]), $\alpha$ is reduced from $\approx 0.64$ (IRR0) to $\alpha \approx 0.46$ (IRR10). Although $\alpha$ can be reduced for doses larger than $1 \times 10^{14}$ cm$^{-2}$, the damage produced by the irradiation considerably diminishes the $J_c$ values at high magnetic fields (Fig. 4*c*). The optimal doses for 6 MeV oxygen irradiation ($\approx 1 \times 10^{14}$ cm$^{-2}$) is 10 times larger than that described in ref. [17] for 3.5 MeV oxygen irradiation. However, as mentioned above, several attempts of irradiation with 3.5 MeV $^{16}$O$^{3+}$ (peak of Bragg within of the GBCO film) showed that, although the in-field of $J_c$ dependence is smoother than for pristine samples, the irradiation reduces considerably the low field $J_c$ values.

In order to analyze the influence of the irradiation in the vortex dynamics, magnetic relaxation measurements at $\mu_0 H = 0.5$ T were performed. This field is within the power-law regime discussed above. The giant flux creep rate observed using magnetic relaxation of the persistent currents is well described by the collective vortex creep model based on the elastic properties of the lattice [25]. This model considers that at small fields and with negligible vortex-vortex interaction (single vortex regime SRV), every single-vortex-line is pinned by the collective action of many weak point-like pinning centers (not observed in thin films with strong pinning centers). The vortex-vortex interactions increase as the magnetic field is incremented and the inter-vortex distance is shortened. Hence, the vortices are collectively trapped in bundles. The normalized creep rate $S = -\delta ln J/\delta ln t$ is given by $S = \frac{T}{U_0 + \mu T \ln(t/t_0)} = \frac{T}{U_0}\left(\frac{J}{J_c}\right)^\mu$ [eq. 1], where $U_0$ is the characteristic energy, $\mu > 0$ is the characteristic glassy exponent and $t_0$ is a characteristic time. Figure 5a shows the $S(T)$ dependences for IRR0, IRR2, IRR4 and IRR10. The initial increment of $S(T)$ at low temperatures can be ascribed to an Anderson–Kim-like mechanism with $S \approx T/U$ (with $U$ the pinning energy). Non-negligible $S$ values are expected at $T = 0$ from quantum creep [25]. At approximately 5 K, all the samples show practically the same $S \approx 0.015$. When the temperature is increased, the $S(T)$ curves present a peak at around 23 K. This is associated with double-kink relaxation due to the presence of correlated pinning. At intermediate temperatures the relaxation displays a minimum with $S(T) \approx$ *constant*, which is related to glassy relaxation. Finally, at high temperatures, the $S$ values are systematically incremented as consequence of thermal smearing of the pinning potential. The main influence of the irradiation or addition of nanoclusters is the reduction of the double-kink peak´s [29]. The expected reduction at the peak is masked by the systematic increment in the $S$ values at the plateau, which can be associated with changes in the µ value that appear in the limit $U_0 \ll \mu T \ln t/t_0$ and it corresponds to $S \approx \frac{1}{\mu \ln t/t_0}$ [22]. To compare the modifications in the vortex dynamics for GBCO CCs irradiated with protons and oxygen ions, the change in the *glassy exponents* µ was analyzed by using the extended Maley's method [30]. The effective activation energy $U_{eff}(J)$ can be experimentally obtained considering the approximation in which the current density decays as $\frac{dJ}{dt} = -\left(\frac{J_c}{\tau}\right)e^{-\frac{U_{eff}(J)}{T}}$. The final equation for the pinning energy is $U_{eff} = -T\left[\ln\left|\frac{dJ}{dt}\right| - C\right]$ (eq. 2) where $C = \ln(J_c/\tau)$ is nominally a constant factor. For an overall analysis it is necessary to consider the function $G(T)$, which results in $U_{eff}(J, T = 0) \approx U_{eff}(J,T)/G(T)$ (eq. 3) [31]. Figure 5b shows the obtained results for IRR0 and IRR10. The inset corresponds to $G(T)$ function. In the limit of $J \ll J_c$ the µ exponent can be estimated as $\Delta \ln U(J) / \Delta \ln J$ [32]. The µ values obtained at intermediate temperatures are summarized in Table I. The µ value is systematically reduced by the doses, for instance $\mu \approx 1.63$ (IRR0) and $\mu \approx 1.45$ (IRR10).

For a better understanding of the vortex dynamics in the pristine and irradiated films, $J_c$ (H) and magnetic relaxation measurements with **H**∥45° were performed. Pinning dominated by nanoparticles similar to $J_c$ (H) dependences are expected for **H** ∥ c and **H**∥45° [9]. Usually for mix pinning landscapes, synergetic combination of twin boundaries and nanoparticles enhances pinning when **H** ∥ c [24]. Figure 6 shows a comparison between $J_c$ (H) for IRR0 and IRR10 at 5 K, 27 K and 40 K. In agreement with the results displayed in Fig. 3 for **H** ∥ c, the irradiation with **H**∥45° enhances the $J_c$ values by improving the in-field dependence. The $J_c$ values are nearly doubling near $\mu_0 H$ = 5 T. Although the $J_c$ (H) dependences can be depicted with a power-law dependence, this behavior is clearly observed for 5 K. At this temperature, α change from ≈ 0.69 (IRR0) to α ≈ 0.47 (IRR10). For 27 K and 40 K, the $J_c$ (H) dependences display a clear change at intermediate fields which cannot be described by only one pinning mechanism or regime and might be related to relaxation by double-kinks [32]. For an overall understanding of the vortex dynamic in the sample for **H**∥45° magnetic relaxation measurements at $\mu_0 H$ = 0.5 T were carried out. Figure 7a shows the S (T) dependences for IRR0, IRR2, and IRR10, which can be described by the same mechanism as for **H** ∥ c. An outstanding feature for S (T) with **H**∥45° is the huge peak associated with relaxation by double-kinks [32,33]. For IRR0 the maximum of the peak is S ≈ 0.04 and it is extended to temperatures above 40 K. The irradiation reduces considerably this relaxation and it is evidenced in the reduction of the absolute value of S at the maximum of the peak. In addition and in agreement with **H** ∥ c, the irradiation increases the absolute S values at the plateau above 40 K. Figure 7b shows the Maley analysis for IRR0 and IRR10, where C = 14 and a G(T) were used (inset Fig. 7b). As well as for **H** ∥ c , the μ value at the collective regime for **H**∥45° is reduced by the irradiation. The μ values change from μ ≈ 1.45 (IRR0) to μ ≈ 1.24 (IRR10).

The results show that similar improvement $J_c$ (H) and changes in the vortex dynamics are produced by proton and oxygen irradiation. The main advantage of oxygen is that the optimal doses are reached in shorter time [17]. Our results are focused on 6 MeV $^{16}O^{3+}$ and on the fact that the peak of Bragg (energy loss) is within the substrate. The optimal dose to improve the in-field dependence of $J_c$ is around $1 \times 10^{14}$ cm$^{-2}$. The glassy exponents μ with **H** ∥ c is reduced by the irradiation from 1.63 (IRR0) to 1.45 (IRR10).

These values are expected for small bundles (3/2-5/2) according to the collective creep model developed for random disorder [25]. It is worth mentioning that pristine 1.3 μm thick YBCO films obtained by MOD display a plateau with $S \approx 0.02$, which is smaller than that observed in the GdBCO tape studied in this work ($S \approx 0.022$). Assuming $ln\ (t/t_0) \approx$ constant, the difference in the $S$ values can be attributed to changes in the vortex bundle size due to different density and morphology of nanoparticles and different density of correlated defects such as TBs. In addition and according to the expectations for pinning by nanoclusters, the irradiation improves the in-field dependence of $J_c$ (H) with $\mathbf{H} \parallel 45°$. At this configuration, the irradiation also suppresses double-kinks and reduces the μ value that dominates the flux creep rates at intermediate temperatures.

The increment in $S$ ($T$) as a function of irradiation fluency, particularly in the intermediate-temperature plateau and for high temperatures, it is common in CCs grown by MOD and co-evaporation [15,16,17,34]. As mentioned above, at the plateau $S \approx \frac{1}{\mu \ln t/t_0}$, the vortex dynamics depends on the μ glassy exponent. The vortex dynamics for pristine CCs usually displays $\mu \approx 3/2$ [14,15,34] which successively decreases with doses. No theoretical predictions have been developed for the glassy exponents in mixed pinning landscapes. Recently, the increment in the S values for irradiated samples was correlated with the different type of defects in mixed pinning landscapes [34]. However, the irradiation in CCs modifies both vortex pinning and the intrinsic superconducting properties of the samples. The detrimental of the superconducting properties is usually evidenced as a gradual suppression of $T_c$ [35]. The optimal doses for irradiation results from a balance between the retention of intrinsic superconducting properties and the enhancement of the vortex pinning. As mentioned before, the $T_c$ reduction produced by the irradiation should have a non-negligible contribution in the absolute value of penetration depth λ [27]. In addition, the anisotropy parameter γ is affected by $T_c$ reduction in the oxygen deficient samples [36]. Local changes in the λ value due to structural disorder should affect properties such as the depairing critical current (maximum theoretical $J_c$ in absence of vortex dissipation) and the elasticity of the vortex lattice [25]. The increment in the vortex fluctuations for irradiated samples is in agreement with the huge suppression of $J_c$ at high temperatures [16]. The contribution of the different type of defects and the suppression of the superconducting properties produced by the damage to the vortex mechanisms require further analysis [34,35]. Future studies should be oriented to understand the influence of the irradiation damage (related to $T_c$ suppression) on properties such as $\lambda$ and $\gamma$. These studies should contribute to understand the role of disorder in the balance between the retention of intrinsic superconducting properties and the enhancement of the vortex pinning.

## 4. Conclusions

In conclusion, we have studied the influence of 6 MeV $^{16}O^{3+}$ irradiation on the critical current densities $J_c$ and flux creep rates in 1.3 μm thick GdBa$_2$Cu$_3$O$_{7-\delta}$ coated conductor produced by co-evaporation. At temperatures below 40 K with $\mathbf{H} \parallel c$ and $\mathbf{H} \parallel 45°$, the critical current density $J_c$ with high magnetic field can be significantly improved by irradiation. For doses of 1x10$^{14}$ cm$^{-2}$ the $J_c$ values at $\mu_0 H$ = 5 T are doubled without affecting significantly the $J_c$ at small fields. Analyzing the flux creep rates as function of the temperature with $\mathbf{H} \parallel c$ and $\mathbf{H} \parallel 45°$, it can be observed that the irradiation suppresses the peak associated with double-kink relaxation and increases the flux creep rates at intermediate and high temperatures. The increment in the flux creep rates at intermediate temperatures can be associated with a reduction in the characteristic glassy exponent $\mu$. While the increment in the flux creep rates at high temperatures may be related with larger vortex fluctuations for irradiated samples. Future studies should be oriented to understand the influence of the irradiation damage (related to $T_c$ suppression) on properties such as $\lambda$ and $\gamma$. An overall understanding of the relationship between pinning and the increment in the vortex fluctuations due to disorder is necessary for the optimization of pinning landscapes in CCs.

**Acknoledgments**

We thank C. Olivares for technical assistance. This work has been partially supported by Agencia Nacional de Promoción Científica y Tecnológica PICT 2015-2171. N. H. is member of the Instituto de Nanociencia y Nanotecnología - CNEA (Argentina).

Table I. Table 1 Summary of oxygen irradiation dose, superconducting critical temperature ($T_c$) and glassy exponenets for **H** // $c$- axis and **H** // 45°. $T_c$ values were obtained from magnetization in 5 Oe with **H**//c-axis after zero field cooling.

| Sample | $T_c$ | μ **H** // $c$- axis | μ **H** // 45° |
| --- | --- | --- | --- |
| Pristine | 92.7 ± 0.2 | 1.63 ± 0.02 | 1.45 ± 0.02 |
| $2 \times 10^{13}$ cm$^{-2}$ | 91.8 ± 0.2 | 1.58 ± 0.02 | 1.47 ± 0.02 |
| $4 \times 10^{13}$ cm$^{-2}$ | 91.0 ± 0.2 | 1.59 ± 0.02 | -- |
| $6 \times 10^{13}$ cm$^{-2}$ | 90.6 ± 0.2 | -- | -- |
| $1 \times 10^{14}$ cm$^{-2}$ | 90.2 ± 0.2 | 1.45 ± 0.02 | 1.24 ± 0.02 |
| $4 \times 10^{14}$ cm$^{-2}$ | 89.8 ± 0.2 | -- | -- |

Figure 1. (a) A TEM image shows typical pinning centers in the GBCO tape. (b) A higher magnification image shows a spherical $Gd_2O_3$ precipitate with diameter around 50 nm.

Figure 2. Temperature dependence of magnetization with $\mu_0 H$ = 0.5 mT in the pristine and 6 MeV Oxygen irradiated films. The magnetization value was normalized by its value at 60 K. The magnetic field was applied parallel to the $c$- axis.

Figure 3. Magnetic field dependence of the critical current densities for IRR0 and for IRR10 at different temperatures. (a) 5 K; (b) 27 K; and (c) 40 K. All measurements were performed with **H**// $c$. The lines between the crossovers are guided by the eye.

Figure 4. (a) α versus oxygen doses obtained from $J_c (H) \propto H^{-\alpha}$. In all cases the values are included for 5 K, 27 K and 40 K. (b) Critical current density $J_c$ versus proton doses at $\mu_0 H \rightarrow 0$ for 5 K, 27 K and 40 K. (c) Critical current density $J_c$ versus proton doses at $\mu_0 H$= 3 T for 5 K, 27 K and 40 K.

Figure 5. (a) Temperature dependence of the creep flux rate ($S$) at $\mu_0 H$ = 0.5 T for IRR0 and oxygen irradiated films (IRR2, IRR4 and IRR10). (b) Maley analysis at $\mu_0 H$ = 0.5 T for IRR0 and IRR10. The inset shows the G($T$) dependences used for the Maley analysis. All measurements were performed with **H** // $c$-axis.

Figure 6. Magnetic field dependence of the critical current density for IRR0 and IRR10 at different temperatures. (a) 5 K; (b) 27 K; and (c) 40 K. All measurements were performed with **H**// 45°. The lines between the crossovers are guided by the eye.

Figure 7. (a) Temperature dependence of the creep flux rate (S) at $\mu_0 H = 0.5$ T for IRR0, IRR2 and IRR10. (b) Maley analysis at $\mu_0 H = 0.5$ T for IRR0 and IRR10. The inset shows the G($T$) dependences used for the Maley analysis. All measurements were performed with **H** // 45°.

Figure 1

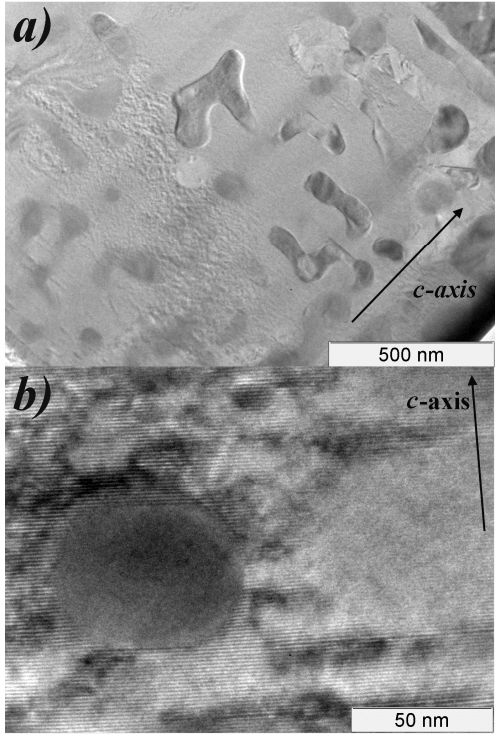

Figure 2

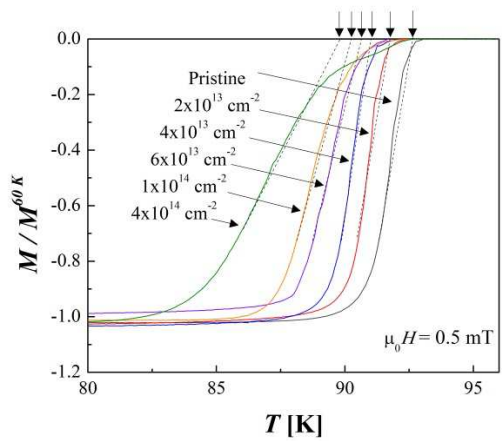

Figure 3

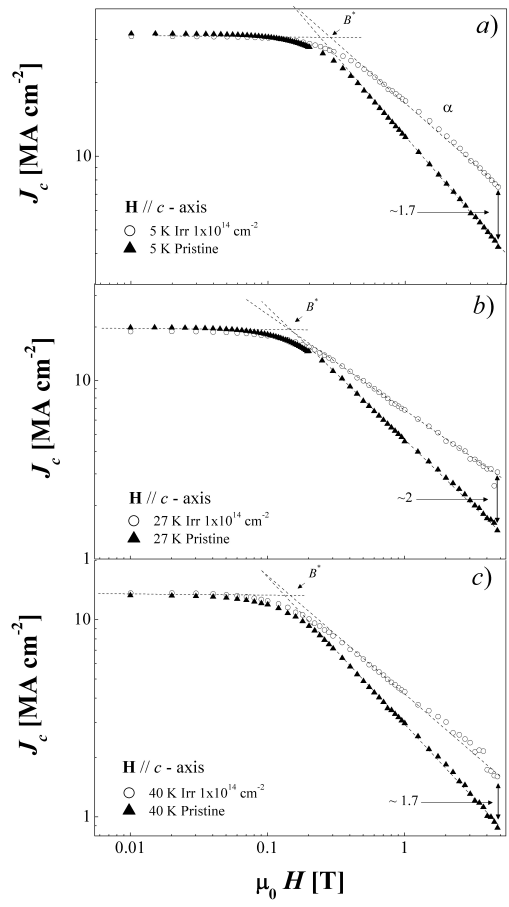

Figure 4

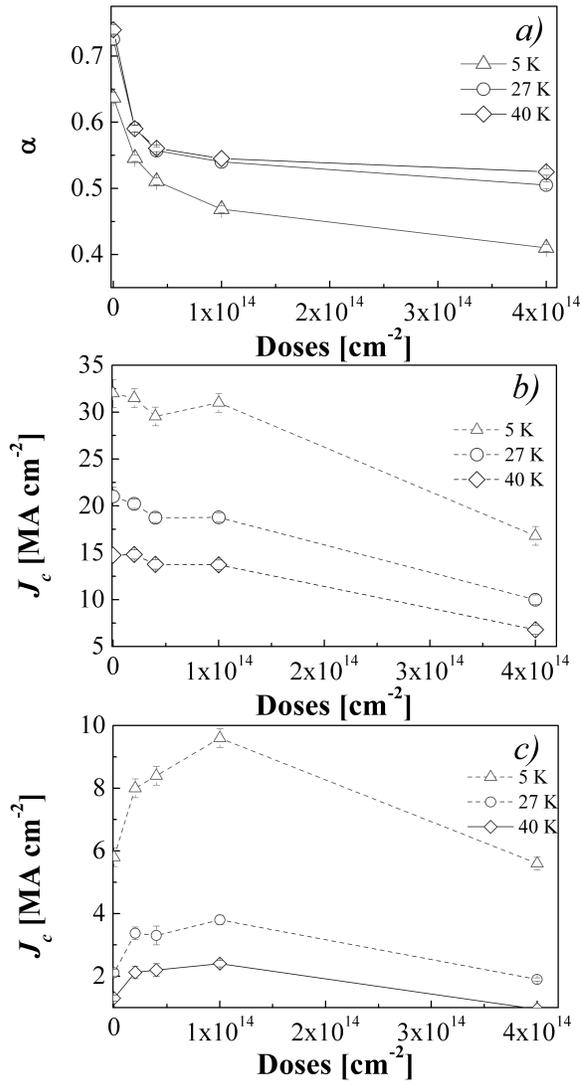

Figure 5

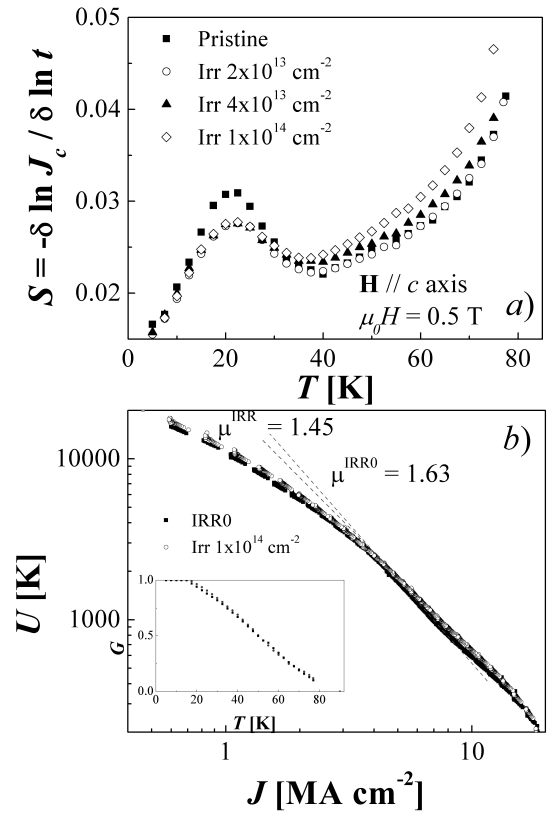

Figure 6

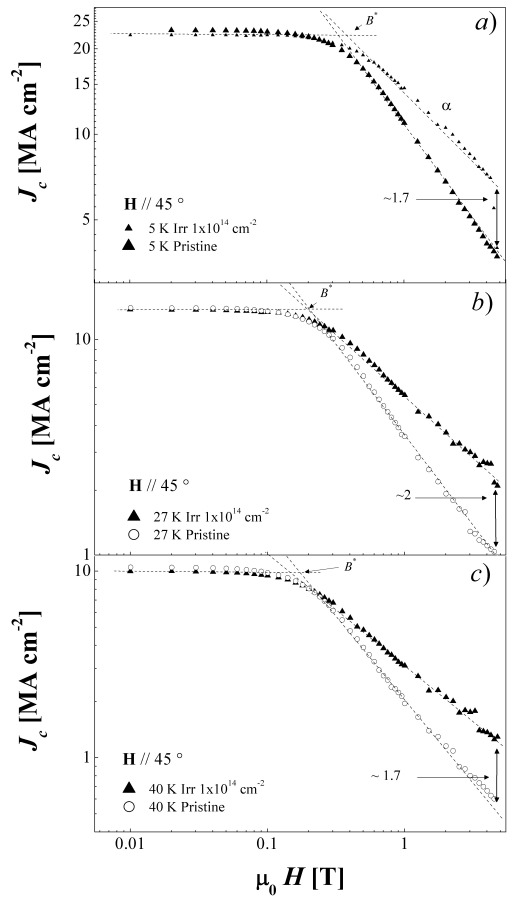

Figure 7

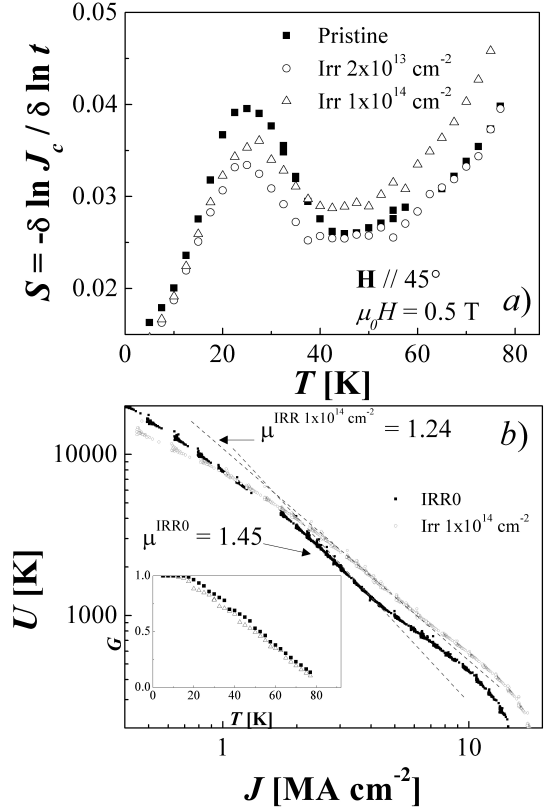

[1] X. Obradors, T. Puig, S. Ricart, M. Coll, J. Gazquez, A. Palau, X. Granados, Growth, nanostructure and vortex pinning in superconducting $YBa_2Cu_3O_7$ thin films based on trifluoroacetate solutions, Supercond. Scie Tech **25** (2012) 123001-123032.
[2] Yuh Shiohara, Takahiro Taneda, Masateru Yoshizumim, Overview of Materials and Power Applications of Coated Conductors Project, Jpn. J. Appl. Phys. 51 (2012) 010007-100722.
[3] M. Heydari Gharahcheshmeh, E. Galstyan, A. Xu, J. Kukunuru, R. Katta, Y. Zhang, G. Majkic, X-F. Li, V. Selvamanickam, Superconducting transition width ($\Delta T_c$) characteristics of 25 mol% Zr-added $(Gd,Y)Ba_2Cu_3O_{7-\delta}$ superconductor tapes with high in-field critical current density at 30 K, Supercond. Sci. Technol. **30** (2017) 015016-015022.
[4] Jae-Hun Lee, Hunju Lee, Jung-Woo Lee, Soon-Mi Choi, Sang-Im Yoo, Seung-Hyun Moon, RCE-DR, a novel process for coated conductor fabrication with high performance, Supercond. Sci. Technol. **27** (2014) 044018-044023.
[5] Carmine Senatore, Matteo Alessandrini, Andrea Lucarelli, Riccardo Tediosi, Davide Uglietti, Yukikazu Iwasa, Progresses and challenges in the development of high-field solenoidal magnets based on RE123 coated conductors, Supercond. Sci. Technol. **27** (2014) 103001-103026.

[6] Sangwon Yoon, Jaemin Kim, Kyekun Cheon, Hunju Lee, Seungyong Hahn, Seung-Hyun Moon, 26 T 35 mm all-$GdBa_2Cu_3O_{7-x}$ multi-width no-insulation superconducting magnet, Supercond. Scie Tech **29** (2016) 04LT04 1-6.

[7] S. R. Foltyn, L.Civale, J. L.Macmanus-Driscoll, Q. X. Jia, B. Maiorov, H. Wang, M. Maley, Materials science challenges for high-temperature superconducting wire, Nat. Mater. **6** (2007) 631-642.

[8] J. L. MacManus-Driscoll, S. R. Foltyn, Q. X. Jia, H. Wang, A. Serquis, L. Civale, B. Maiorov, M. E. Hawley, M. P. Maley, D. E. Peterson, Strongly enhanced current densities in superconducting coated conductors of $YBa_2Cu_3O_{7-x}$ + $BaZrO_3$, Nat. Mat. **3** (2004) 439-443.

[9] N. Haberkorn, M. Miura, J. Baca, B. Maiorov, I. Usov, P. Dowden, S. R. Foltyn, T. G. Holesinger, J. O. Willis, K. R. Marken, T. Izumi, Y. Shiohara, L. Civale, High-temperature change of the creep rate in $YBa_2Cu_3O_7$ films with different pinning landscapes, Phys. Rev B **85** (2012) 174504-174510.

[10] J. Albrecht, M. Djupmyr, S. Brück, Universal temperature scaling of flux line pinning in high-temperature superconducting thin films, J. Phys.: Condens. Matter 19 (2007) Article 216211.

[11] Kaname Matsumoto, Paolo Mele, Artificial pinning center technology to enhance vortex pinning in YBCO coated conductors, Supercond. Sci. Technol. **23** (2010) 014001-014012.

[12] V. Selvamanickam, M. Heydari Gharahcheshmeh, A. Xu, Y. Zhang, E. Galstyan, Critical current density above 15MAcm$^{-2}$ at 30K, 3T in 2.2μm thick heavily-doped $(Gd,Y)Ba_2Cu_3O_x$ superconductor tapes, Supercond. Sci. Technol. **28** (2015) 072002-072006.

[13] M. Miura, B. Maiorov, J. O. Willis, T. Kato, M. Sato, T. Izumi, Y. Shiohara, L. Civale, The effects of density and size of $BaMO_3$ (M=Zr, Nb, Sn) nanoparticles on the vortex glassy and liquid phase in $(Y,Gd)Ba_2Cu_3O_y$ coated conductors, Supercond. Sci. Technol. **26** (2013) 035008-025014.

[14] N. Haberkorn, Y. Coulter, A. M. Condó, P. Granell, F. Golmar, H. S. Ha, S. H. Moon, Vortex creep and critical current densities $J_c$ in a 2 μm thick $SmBa_2Cu_3O_{7-d}$ coated conductor with mixed pinning centers grown by co-evaporation, Supercond Scie Tech **29** (2016 ) 075011-075017.

[15] Y. Jia *et al.,* Doubling the critical current density of high temperature superconducting coated conductors through proton irradiation, Appl. Phys. Lett. **103** (2013) Article 122601 .

[16] N. Haberkorn, Jeehoon Kim, S. Suárez, Jae-Hun Lee, S. H. Moon, Influence of random point defects introduced by proton irradiation on the flux creep rates and magnetic field dependence of the critical current density $J_c$ of co-evaporated $GdBa_2Cu_3O_{7-\delta}$ coated conductors, Supercond. Sci. Technol. **28** (2015) 125007-125012.

[17] M. Leroux *et al.,* Rapid Doubling of the Critical Current of $YBa_2Cu_3O_{7-\delta}$ Coated Conductors for Viable High-Speed Industrial Processing. Appl. Phys. Lett **107** (2015) Article 192601.

[18] I. A. Sadovskyy et *al.,* Toward Superconducting Critical Current by Design*,* Adv. Mater. **28** (2016) 4593-4600.

[19] Wai-Kwong Kwok, Ulrich Welp, Andreas Glatz, Alexei E. Koshelev, Karen J. Kihlstrom, George W. Crabtree, Vortices in high-performance high-temperature superconductors, Rep. Prog. Phys **79** (2016) 116501-116539.

[20] J. L. MacManus-Driscoll, M. Bianchetti, A. Kursumovic, G. Kim, W. Jo, H. Wang, J. H. Lee, G. W. Hong, S. H. Moon, Strong pinning in very fast grown reactive co-evaporated $GdBa_2Cu_3O_7$ coated conductors, APL Materials **2** (2014) 086103-086110.

[21] N. Haberkorn *et al.*, Influence of random point defects introduced by proton irradiation on critical current density and vortex dynamics of $Ba(Fe_{0.925}Co_{0.075})_2As_2$ single crystals Phys. Rev B **85** (2012) 014522-014528.

[22] Y. Yeshurun, A. P. Malozemoff, A. Shaulov, Magnetic relaxation in high-temperature superconductors, Rev. Mod. Phys. **68** (1996) 911-949.

[23] J. F. Ziegler, J. P. Biersack, U. Littmark, 1985 The Stopping and Range of Ions in Solids (New York: Pergamon).

[24] M. Miura, B. Maiorov, S. A. Baily, N. Haberkorn, J. O. Willis, K. Marken, T. Izumi, Y. Shiohara, L. Civale, Mixed pinning landscape in nanoparticle-introduced YGdBa2Cu3Oy films grown by metal organic deposition, Phys. Rev B **83** (2011) 184519-184526.

[25] G. Blatter, M. V. Feigelman, V. B. Geshkenbein, A. I. Larkin, V. M. Vinokur, Vortices in high-temperature superconductors, Rev. Mod. Phys. **66** (1994) 1125-1388.

[26] N. Basov, A. V. Puchkov, R. A. Hughes, T. Strach, J. Preston, T. Timusk, D. A. Bonn, R. Liang, W. N. Hardy, Disorder and superconducting-state conductivity of single crystals of $YBa_2Cu_3O_{6.95}$, Phys. Rev. B **49** (1994) 12165-12169.

[27] Y. J. Uemura *et al.*, Universal Correlations between $T_c$ and $n_s/m^*$ (Carrier Density over Effective Mass) in High-Tc Cuprate Superconductors Phys. Rev Lett **62** (1989) 2317-2320.

[28] C. J. van der Beek, M. Konczykowski, A. Abal'oshev, I. Abal'osheva, P. Gierlowski, S. J. Lewandowski, M. V. Indenbom, S. Barbanera, Strong pinning in high-temperature superconducting films, Phys. Rev. B **66** (2002) 024523-024532.

[29] B. Maiorov, S. A. Baily, H. Zhou, O. Ugurlu, J. A. Kennison, P. C. Dowden, T. G. Holesinger, S. R. Foltyn, L. Civale, Synergetic combination of different types of defect to optimize pinning landscape using $BaZrO_3$-doped $YBa_2Cu_3O_7$, Nat. Mater. **8** (2009) 398-404.

[30] M. P. Maley, J. O. Willis, H. Lessure, M. E. McHenry, Dependence of flux-creep activation energy upon current density in grain-aligned $YBa_2Cu_3O_7$, Phys. Rev. B **42** (1990) 2639-2642.

[31] J. G. Ossandon, J. R. Thompson, D. K. Christen, B. C. Sales, Y. Sun, K. W. Lay, Flux-creep studies of vortex pinning in an aligned $YBa_2Cu_3O_7$ superconductor with oxygen deficiencies $\delta \leq 0.2$, Phys. Rev. B **46** (1992) 3050-3058.

[32] J. R. Thompson, L. Krusin-Elbaum, L. Civale, G. Blatter, C. Field, Superfast Vortex Creep in $YBa_2Cu_3O_{7-d}$ Crystals with Columnar Defects: Evidence for Variable-Range Vortex Hopping, Phys. Rev Lett. **78** (1997) 3181-3184.

[33] L. Civale, Vortex pinning and creep in high-temperature superconductors with columnar defects, Supercond. Sci. Technol. **10** (1997) A11-A29.

[34] Eley S, Leroux M, Rupich M W, Miller D J, Sheng H, Niraula P M, Kayani A, Welp U, Kwok W-K and L Civale, Decoupling and tuning competing effects of different types of defects on flux creep in irradiated $YBa2Cu3O_{7-d}$ coated conductors, Supercond. Sci. Technol. **30** (2017) 015010.

[35] N. Haberkorn, J. Guimpel, S. Suárez, H. Troiani, P. Granell, F. Golmar, Jaehun Lee, S-H Moon and Hunju Lee, Strong influence of the oxygen stoichiometry on the vortex bundle size and critical current densities $J_c$ of $GdBa_2Cu_3O_x$ coated conductors grown by co-evaporation, Supercond. Sci. Technol. **30** (2017) 95009.

[36] T.R. Chien, W.R. Datars, B.W. Veal, A.P. Paulikas, P. Kostic, Chun Gu, Y. Jiang, Dimensional crossover and oxygen deficiency in $YBa_2Cu_3O_\chi$ single crystals, Physica C **229** (1994) 273-279.